# Runout of Liquefaction-Induced Tailings Dam Failure: Influence of Earthquake Motions and Residual Strength

Brent Sordo, Ph.D.[1], Ellen Rathje, Ph.D.[2], Krishna Kumar, Ph.D.[3]

[1]Former Ph.D. Candidate, Department of Civil, Architectural, and Environmental Engineering, The University of Texas at Austin, 301 E Dean Keeton St, Austin, TX 78712; E-mail: bsordo@utexas.edu
[2]Professor, Department of Civil, Architectural, and Environmental Engineering, The University of Texas at Austin, 301 E Dean Keeton St, Austin, TX 78712; E-mail: e.rathje@mail.utexas.edu
[3]Assistant Professor, Department of Civil, Architectural, and Environmental Engineering, The University of Texas at Austin, 301 E Dean Keeton St, Austin, TX 78712; E-mail: krishnak@utexas.edu

## ABSTRACT

This study utilizes a hybrid Finite Element Method (FEM) and Material Point Method (MPM) to investigate the runout of liquefaction-induced flow slide failures. The key inputs to this analysis are the earthquake ground motion, which induces liquefaction, and the post-liquefaction residual strength. The influence of these factors on runout is evaluated by subjecting a model of a tailings dam to thirty different earthquake motions and by assigning different values of post-liquefaction residual strength. Ground motions with larger peak ground accelerations (PGA) generate liquefaction to larger depths, thus mobilizing a greater mass of material and resulting in a flow slide with greater runout. However, different ground motions with the same PGA yield significant variations in the depth of liquefaction, indicating that other ground motion characteristics (e.g., frequency content) also exert significant influence over the initiation of liquefaction. Ground motion characteristics of peak ground velocity (PGV) and Modified Acceleration Spectrum Intensity (MASI) show a strong correlation to the induced depth of liquefaction because they capture both the intensity and frequency content of the earthquake motion. The computed runout is directly related to the depth of liquefaction induced by the earthquake motion. For dam geometry analyzed, measurable runout occurs when liquefaction extends to 10 m depth and the runout is maximized when liquefaction extends to about 18 m. Strain-softening of the residual strength of the liquefied tailings during runout is shown to substantially increase the runout distance of the flow slide, highlighting the need for additional research to better characterize the appropriate strength of liquefied materials during flow failures.

**INTRODUCTION**

Earthquake-induced liquefaction has the potential to trigger massive flow slides that may destroy infrastructure, threaten downstream communities, and damage local ecosystems [1,2]. Liquefaction causes sudden strength loss and can result in the failure of natural or anthropogenic slopes, releasing flow slides that extend for kilometers [3]. Structures particularly susceptible to liquefaction-induced failures during seismic events include hydraulic fill dams (e.g., Lower San Fernando dam) [4], tailings dams (e.g., Kayakari dam) [5,6], and other large embankments composed of loose, saturated sands. Given the catastrophic potential of such failures, it is crucial for earthquake engineers to develop robust methods for assessing liquefaction susceptibility, predicting potential flow slide runout, and designing mitigation strategies to enhance the resilience of critical infrastructure in seismically active regions.

Analyzing liquefaction-induced failures involves two critical components: predicting the onset of liquefaction and estimating the subsequent runout. Accurately assessing the triggering of liquefaction is fundamental to understanding these hazards. The strength reduction due to liquefaction is key to initiating these failures [7,8], making the estimation of excess pore pressure development crucial. While the simplified stress-based empirical approach, which primarily uses peak ground acceleration (PGA), remains the standard for predicting seismically induced liquefaction [9], research has shown that other ground motion characteristics, such as frequency content and duration, significantly influence liquefaction potential [10,11]. An alternative approach involves finite element analysis with advanced constitutive models for liquefiable sand, which can account for all ground motion characteristics in predicting excess pore pressure generation, liquefaction triggering, and resulting instability [7,12,13]. FEM offers a more comprehensive understanding of liquefaction onset, crucial for subsequent runout analyses and the development of effective mitigation strategies.

The extent of runout of these flow slides is complex and depends on many factors such as dam design, downstream ground surface geometry, groundwater depth, the extent of liquefaction caused by the earthquake, and the residual strength of the liquefied materials [14,15]. Predicting these runouts is important to hazard mitigation efforts [16]. Consequently, the use of numerical methods (e.g., the Material Point Method, MPM, [17,18]) to predict flow slide runouts is a common topic of research (e.g., [19]), although standardized procedures have not yet been established [16,20].

This paper investigates the sensitivity of the runout of liquefaction-induced tailings dam failures to earthquake ground motion characteristics and post-liquefaction residual strength. The study employs a sequential hybrid FEM-MPM method [21,22] to analyze a tailings dam subjected to various earthquake motions. The geometry of the tailings dam is based on the Mochikoshi tailings dam that failed during an earthquake in 1978. The FEM phase simulates liquefaction triggering and failure initiation, while the MPM phase models the subsequent runout. The analysis evaluates how different ground motion parameters influence the extent of liquefaction and assesses the impact of residual strength characterization on runout distances. Ten different acceleration-time history recordings are selected, and each is scaled to three different PGAs (0.2, 0.3, and 0.4 g) for a total of thirty ground motions. Each of these analyses results in a unique failure initiation scenario which is transferred to MPM. Four characterizations of residual strength of the liquefied tailings in the MPM phase are considered ($c_{soft}/S_r$ = 1.0, 0.75, 0.50, and 0.25), yielding a total of 120 runout analyses. By examining the relationship between these variables and the flow slide runout, this research aims to improve our understanding of the key factors controlling liquefaction-induced flow slides and to contribute to more accurate risk assessments for tailings dams in seismically active regions.

**HYBRID FEM-MPM MODEL OF MOCHIKOSHI TAILINGS DAM**

This study applies the hybrid FEM-MPM approach to the Mochikoshi tailings dam No. 1, a 30-m high structure in Japan that failed during the 1978 Izu-Ohshima-Kinkai earthquake (Mw = 7.0). Seismic loading and liquefaction-induced strength reduction of the tailings caused the failure, sending a large mass of liquefied material flowing approximately 800 meters downstream [23,24]. Figure 1 shows the tailings dam geometry, which includes lower and upper embankment dams, tailings, and foundation soils. Sordo et al. [22] recently analyzed this failure using a hybrid FEM-MPM approach using a single input ground motion time history from the 1971 San Fernando (Mw = 6.6) earthquake. This analysis is summarized below.

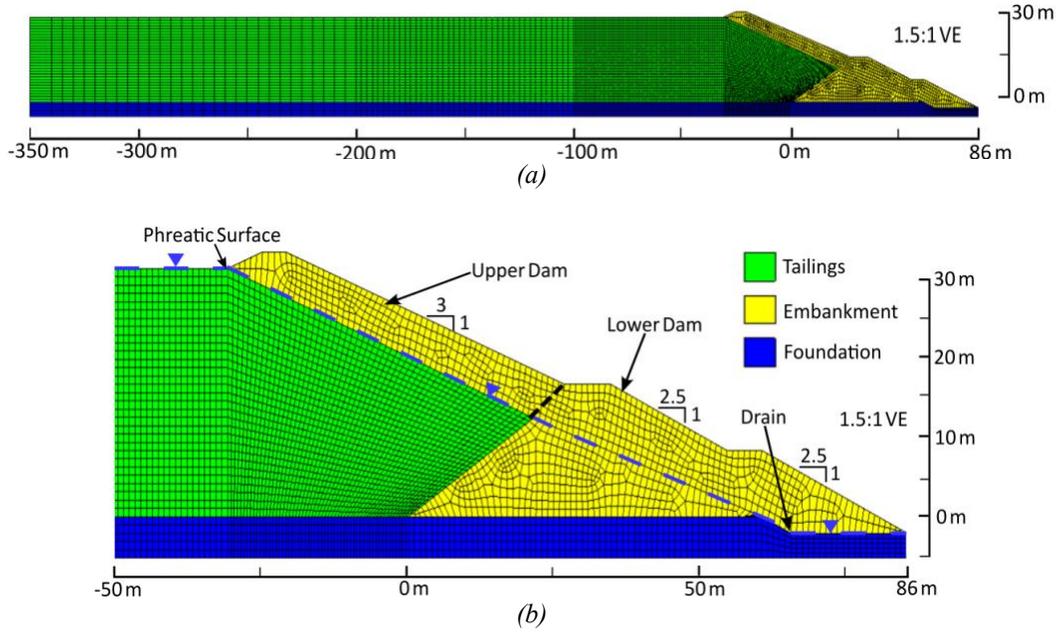

*Figure 1: (a) Full view of the mesh of the FEM phase of the Mochikoshi hybrid model and (b) focused view of the unstructured mesh of the dam.*

A hybrid FEM-MPM analysis consists of three phases: an initiation phase modeled in FEM, a transfer phase that shifts the model from FEM to MPM, and a runout phase modeled in MPM. For a more detailed description of the general hybrid FEM-MPM procedure, see our previous publication [21]. The FEM phase of the hybrid Mochikoshi analysis was conducted via the FEM program OpenSees [25]. The model features free field columns at the left and right boundaries and a Lysmer-Kuhlemeyer dashpot [26] at the base. The model was divided into three materials: embankment, tailings, and foundation (Figure 1). Each of these materials was modeled via the PM4Sand constitutive model, which was developed to characterize liquefiable soils in earthquake applications [27]. Mass densities, permeabilities, and void ratios were assigned to each of the materials based on the documentation of the failure (e.g., [23,24]). The cyclic resistance (i.e., cyclic resistance ratio, CRR) and post-liquefaction residual strength ($S_r$) of the tailings were assigned based on a clean sand, stress-corrected blow count ($N_{1,60\text{-cs}}$) of 6, a value estimated from data provided by Ishihara [24] and Byrne and Seid-Karbasi [28]. The PM4Sand model for the tailings was calibrated to the PM4Sand default CRR for $D_R$ of 35% sand [27], which is consistent with $N_{1,60\text{-cs}} = 6$, and the stress-dependent $S_r$ values tabulated in Table 1, which are derived from the $S_r/\sigma'_v$ empirical relationship of Weber [29]. The calibration process involves utilizing undrained monotonic and cyclic shear tests to iteratively adjust $S_r$ and CRR until the desired values of both are reached; it is described in detail in our previous publication [22]. The embankment and foundation materials were assigned the default properties of a medium sand ($D_R$ = 55%) and a dense sand ($D_R$ = 75%), respectively. The tailings behave undrained under the seismic conditions, but embankment is either dry (due to a drain at the bottom of the dam; Figure 1b) or behaves drained (due to higher permeability than the tailings). Table 2 shows the material parameters of all three materials. The earthquake was simulated by applying the ground motion to the dashpot, triggering the failure. The FEM phase captures the key features of the failure initiation – the development of excess pore water pressures, liquefaction, and a shear surface – before the simulation is transferred into the MPM phase.

*Table 1: Undrained shear strengths and corresponding consolidation stresses from Weber (2015) to which constitutive model of tailings in FEM phase is calibrated.*

| $\sigma'_{vo}$ (kPa) | $s_r$ (kPa) |
|---|---|
| 50 | 4.3 |
| 100 | 6.5 |
| 250 | 11.1 |

*Table 2: Material properties in the FEM phase of the Mochikoshi hybrid model.*

| Material | Dr | $G_0$ | $h_{po}$ | Q | R | $\gamma$ (kN/m³) | e | $k_x$ (m/s) | $k_y$ (m/s) |
|---|---|---|---|---|---|---|---|---|---|
| Tailings | 0.35 | 476 | 2.70 | 11.998 | 3.75 | 18.3 | 0.99 | 7.1E-06 | 7.1E-09 |
| Embankment | 0.55 | 677 | 0.40 | 10 | 1.50 | 16.6 | 1.20 | 1.0E-06 | 1.0E-06 |
| Foundation | 0.75 | 890 | 0.63 | 10 | 1.50 | 18.6 | 0.80 | 1.0E-05 | 1.0E-05 |

Figure 2a shows the deviatoric strains in the FEM mesh at $t$ = 17.0 s into the earthquake motion. The deviatoric strain contours display the formation of two distinct failure surfaces, a key characteristic that developed as liquefaction extended to deeper depths. The development of excess pore pressures is quantified through the pore pressure ratio, $r_u$, defined as:

$$r_u = \frac{\Delta u}{\sigma'_{vo}}$$ (1)

where $\Delta u$ is the excess pore pressure and $\sigma'_{vo}$ is the initial effective vertical stress. Large values of $r_u$ indicate liquefaction. Figure 2b shows contours of $r_u$ at $t$ = 17.0 s, with values of $r_u$ greater than 0.7 through much of the tailings. Accurately capturing the extent of liquefaction in the FEM phase is key, as the associated strength reduction governs the development of the failure mechanism. It is important that the earthquake-induced liquefaction is fully developed during the FEM phase before the transfer to MPM, as the ground motion is not applied in the MPM phase.

The FEM to MPM transfer involves mapping the geometry, state variables and material properties of each finite element into MPM material points using a Python script (available at [30]). This transfer occurs at a user-specified time, selected to ensure the failure mechanism is fully developed but before the FEM mesh becomes excessively distorted. For liquefaction failures, the FEM-MPM transfer phase also involves identifying which material points represent liquefied soil and assigning them appropriate post-liquefaction residual strengths. The extent of liquefaction is defined based on the $r_u$ values (Figure 2b), with locations achieving $r_u > 0.7$, and those above them, assigned as liquefied. This threshold was selected because embankment materials with $r_u > 0.7$ have been shown to exhibit the potential for cyclic mobility (i.e., the potential for large strains) [31] due to the presence of high static shear stresses.

The MPM phase is conducted via the CB-Geo MPM code [32] and simulates the runout of the dam failure using material points that move through a structured background grid independent of the distorted FEM mesh. The material points, which have received their initial conditions from the FEM phase via the transfer algorithm, are allowed to displace until the runout mass has reached its final geometry. Because the development of liquefaction is no longer simulated in the MPM phase, a simple Mohr-Coulomb constitutive model is used in the MPM phase. To account for the stress-dependence of $S_r$, the material points in the liquified zone are divided into layers representing different values of $S_r$ based on their $\sigma'_{v0}$ (Figure 2c; Table 3) according to the relationship of Weber [29], and each layer is assigned $\phi = 0$ with $c = S_r$. Some analyses also considered further strain softening of the residual strength by assigning $c_{soft} = 0.5 \cdot S_r$. The unliquefied tailings and embankment are assigned strength parameters based on the strength characterization of Byrne and Seid-Karbasi [28].

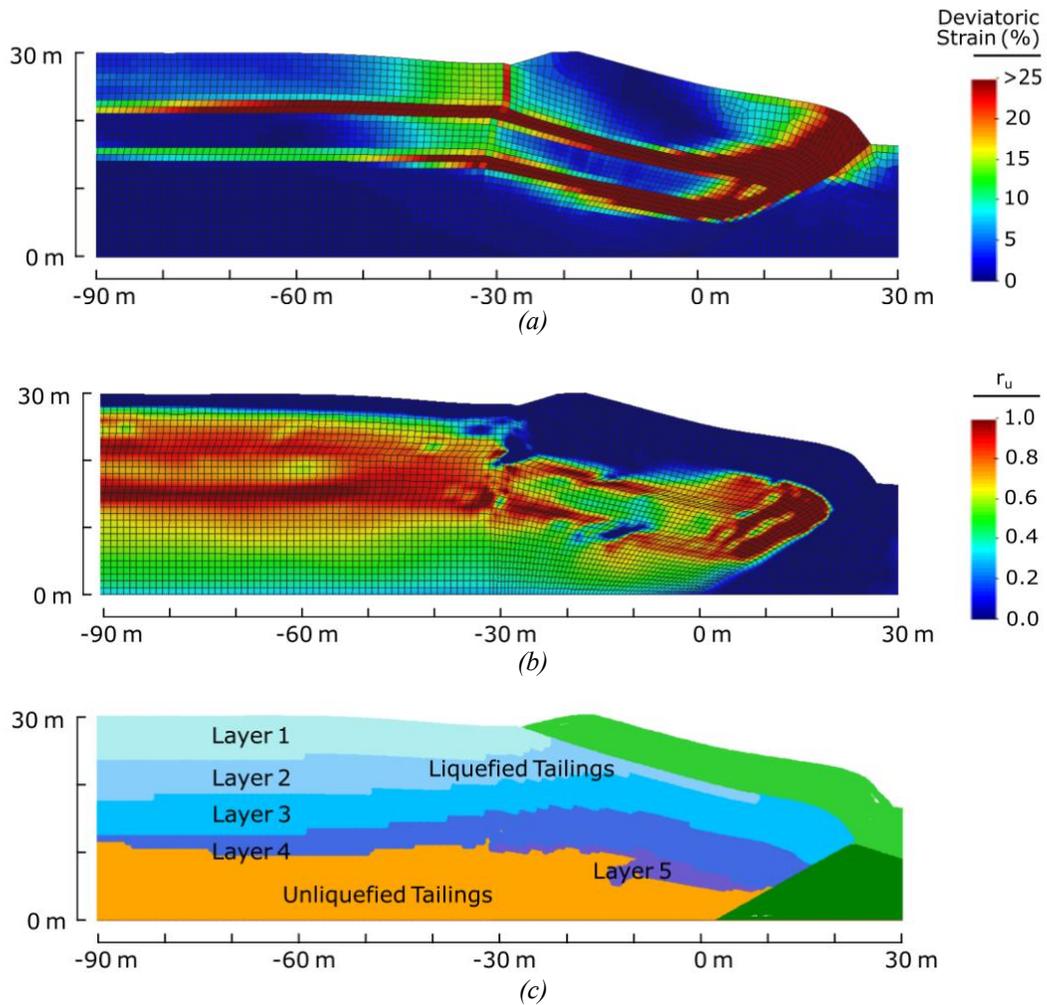

*Figure 2: Results from FEM and transfer phases of Mochikoshi failure: (a) deviatoric strain contours of the FEM phase at t = 17.0 s, (b) $r_u$ contours of FEM phase at t = 17.0 s, and (c) distribution of liquefaction transferred to the MPM phase. Reproduced after Sordo et al. [22].*

*Table 3: Residual shear strengths and corresponding initial effective vertical stress ranges from Weber (2015) to which constitutive models of liquefied tailings in MPM phase are calibrated.*

| Material | $\phi$ (°) | $c$ (kPa) |
|---|---|---|
| Embankment | | |
| Saturated | 35 | 25 |
| Unsaturated | 35 | 25 |
| Unliquefied Tailings | 30 | 1 |
| Liquefied Tailings | | |
| Layer 1 | 0 | 3 |
| Layer 2 | 0 | 5.6 |
| Layer 3 | 0 | 7.4 |
| Layer 4 | 0 | 9.1 |
| Layer 5 | 0 | 10.6 |

Figure 3 shows the final runout geometry from the MPM analyses with the FEM to MPM transfer occurring at $t = 17.0$ s. The particles are allowed to run out over a horizontal frictional boundary ($\mu = 0.70$), and the left free field column is replaced by a vertical roller boundary. Runout distance is defined as the distance of the fifth-farthest material point from the toe of the dam, to avoid the influence of unstable material points. The analysis reveals that while the upper dam completely fails, the lower dam remains largely intact. The resulting runout extends approximately 35 m downstream from the toe of the dam when the residual strength does not soften ($c_{soft} = 1.0 \cdot S_r$), but the runout distance increases to 215 m when strain softening is included in the liquefied tailings ($c_{soft} = 0.5 \cdot S_r$).

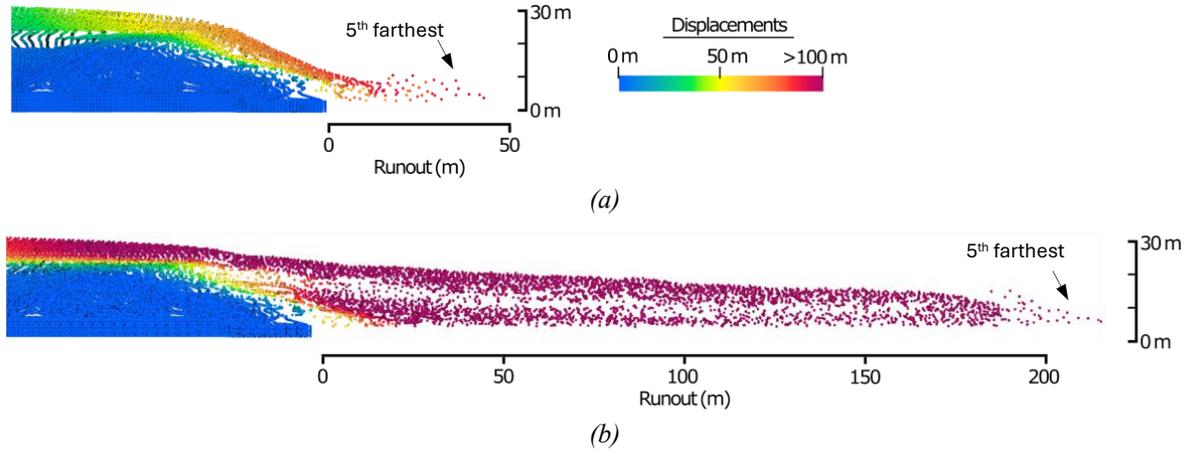

*Figure 3: Runouts from the MPM phase of $t_T$ = 17.0 s transfer of the Mochikoshi failure when (a) strain softening is excluded in the liquefied tailings ($c_{soft} = 1.0 \cdot S_r$) and (b) when strain softening is included in the liquefied tailings ($c_{soft} = 0.5 \cdot S_r$). Reproduced after Sordo et al. [22]. Note that these images are cropped on the left side. All of the model area is included in the transfer with the exception of the free field columns.*

Our previous analysis focused on identifying appropriate FEM to MPM transfer times for failures caused by earthquake-induced liquefaction. As a result, they considered only one input ground motion and two characterizations of the residual strength. For the sensitivity analyses performed in this study, the same Mochikoshi dam model is subjected to thirty different ground motions and considers four degrees of strain-softening in the liquefied tailings. These analyses will provide insights into the influence of earthquake ground motions and residual strength characterization on the triggering and runout of liquefaction-induced flow slides. For all the analyses performed in this study, the guidance of our previous analyses [21,22] were used to select an appropriate transfer time based on liquefaction, deviatoric strain, kinetic energy, and mesh distortion criteria.

## INPUT EARTHQUAKE GROUND MOTIONS

Ten earthquake acceleration-time history recordings (Table 4) were selected from the NGA-West2 database [33] to represent a wide range of earthquake magnitudes, significant durations ($D_{5-95}$, defined as the time between 5 and 95% of the buildup of the Arias Intensity; [34]), and frequency contents ($T_m$, mean period; [35]). Figure 4a shows the waveforms of each ground motion, and Figure 4b displays the normalized pseudo-acceleration response spectra (PSA / PGA). The differences in the duration of strong shaking observable between the ground motions in Figure 4a is reflected in the differences in $D_{5-95}$ (Table 4), which range from 9.5 s (i.e., Loma Prieta) to 27.4 s (i.e., Hector Mine). The differences in frequency content can also be observed in Figure 4a. The Corinth and Hector Mine ground motions have noticeably lower frequency oscillations, while the Loma Prieta and Irpinia ground motions have noticeably higher frequency oscillations. The differences in frequency content are more clearly visible in the response spectra (Figure 4b). The peaks of the response spectra all occur at relatively similar periods between T = 0.1 - 0.5 s, but the spectra exhibit significantly different responses at T > 0.4 s. At T = 2 s, there is more than an order of magnitude difference between the response spectral values. The largest $T_m$ motion (i.e., Hector Mine) yields the largest spectral ordinates at long periods, while the smallest $T_m$ motion (i.e., Loma Prieta) yields the smallest.

*Table 4: Input Earthquake Ground Motion Characteristics*

| Earthquake | $M_w$ | Station | $R_{rup}$ (km) | $V_{S30}$ (m/s) | $D_{5-95}$ (s) | $T_m$ (s) |
|---|---|---|---|---|---|---|
| 1992 Big Bear | 6.5 | Rancho Cucamonga – Deer Can (090) | 60 | 509 | 25.4 | 0.40 |
| 1992 Cape Mendocino | 7.0 | Eureka – Myrtle & West (000) | 42 | 337 | 20.6 | 0.75 |
| 1981 Gulf of Corinth | 6.6 | Corinth, Greece (L) | 10 | 361 | 15.4 | 0.70 |
| 1999 Hector Mine | 7.1 | Amboy CSMIP Station (090) | 43 | 383 | 27.4 | 0.77 |
| 1980 Irpinia | 6.9 | Brienza (000) | 23 | 561 | 13.2 | 0.31 |
| 1992 Landers | 7.3 | Yermo Fire Station (000) | 24 | 354 | 20.1 | 0.66 |
| 1989 Loma Prieta | 7.0 | UCSC Lick Observatory (090) | 18 | 714 | 9.5 | 0.22 |
| 2019 Ridgecrest | 7.1 | CIT China Lake Station (090) | 7 | 1227 | 10.4 | 0.39 |
| 1971 San Fernando | 6.6 | CIT Athaneum Library (090) | 25 | 415 | 14.5 | 0.56 |
| 1978 Tabas | 7.4 | Dayhook (L) | 14 | 472 | 11.3 | 0.33 |

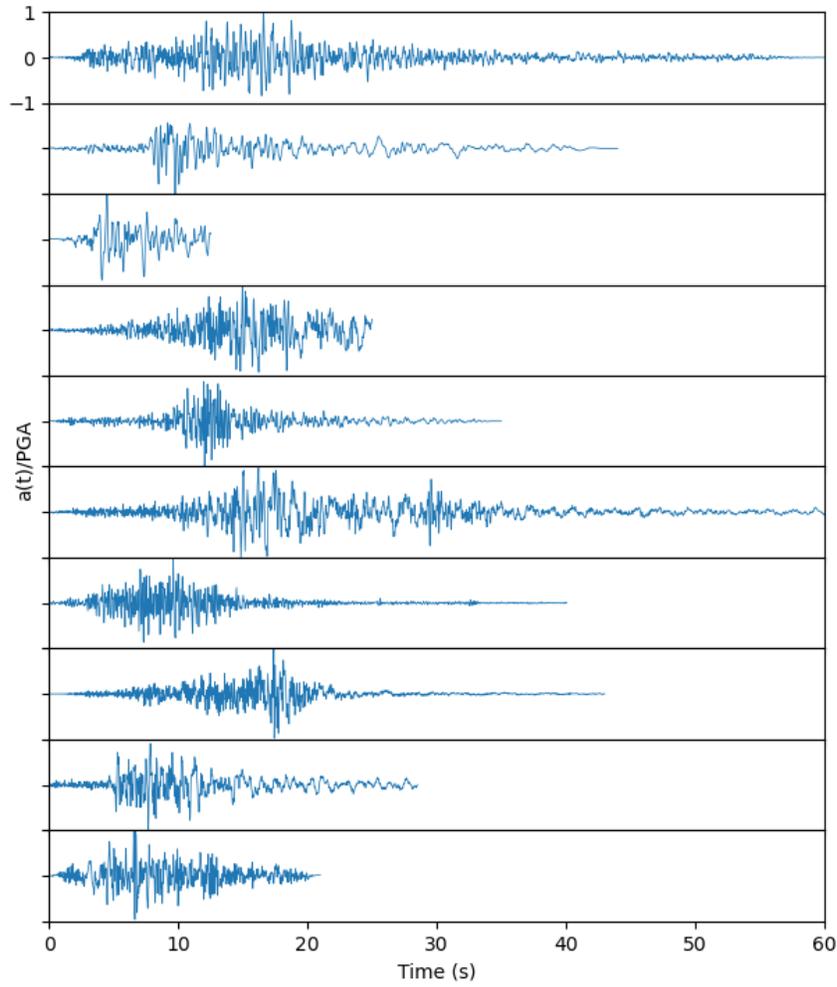

*(a)*

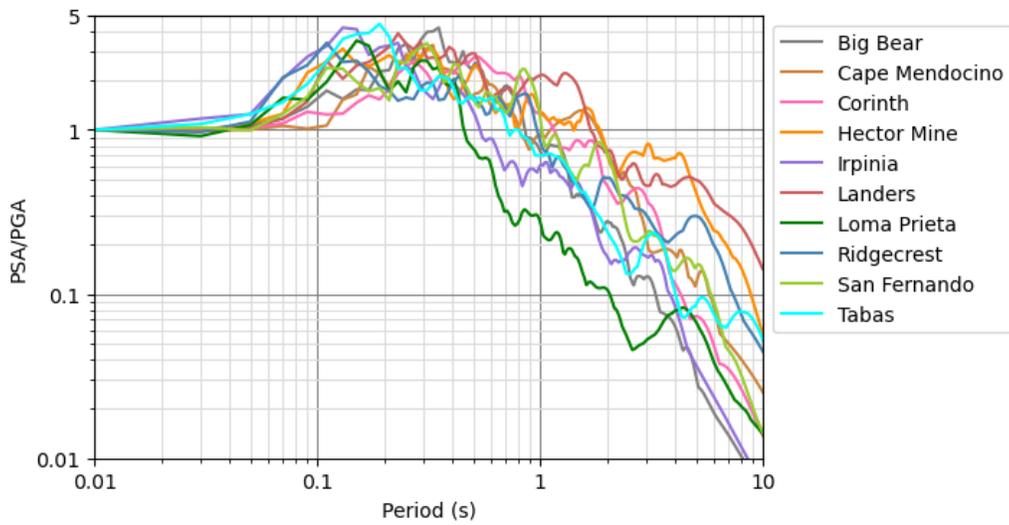

*(b)*

*Figure 4: (a) Acceleration-time histories and (b) normalized pseudo-acceleration response spectra of the input ground motions.*

To evaluate the influence of PGA on the FEM-MPM results, each ground motion in Figure 4a is scaled to PGA values of 0.2, 0.3, and 0.4 g, yielding a total of 30 ground motions. While PGA is the standard earthquake intensity measure (IM) for predicting earthquake-induced liquefaction [9], research has shown that other IMs may show stronger correlations [10,11]. Therefore, this study also considers the peak ground velocity (PGV) and modified acceleration spectrum intensity (MASI) of each ground motion to investigate their relationship to the development of liquefaction. PGV is the peak in the velocity-time history, and MASI is the integral of the pseudo-acceleration response spectrum between periods of 0.1 and 1.5 s [11]. PGV and MASI have been found to be superior predictors of liquefaction because they have the advantage of implicitly accounting for both PGA and frequency content. Figure 5 plots the PGV and MASI of the scaled input motions as a function of the scaled PGA. While these IMs have a positive correlation to PGA, the values of PGV and MASI of the ground motions vary by a factor of 3 at each specific PGA, reflecting the difference in frequency contents between them. For example, the Loma Prieta ground motion has the smallest $T_m$ and also has the smallest PGV and MASI ($T_m$ = 0.22 s, PGV = 16.0 cm/s and MASI = 3.35 m/s at PGA = 0.3 g), while the Hector Mine ground motion has the largest values of $T_m$ and among the largest PGV and MASI ($T_m$ = 0.77 s, PGV = 45.6 cm/s, and MASI = 6.77 m/s at PGA = 0.3 g).

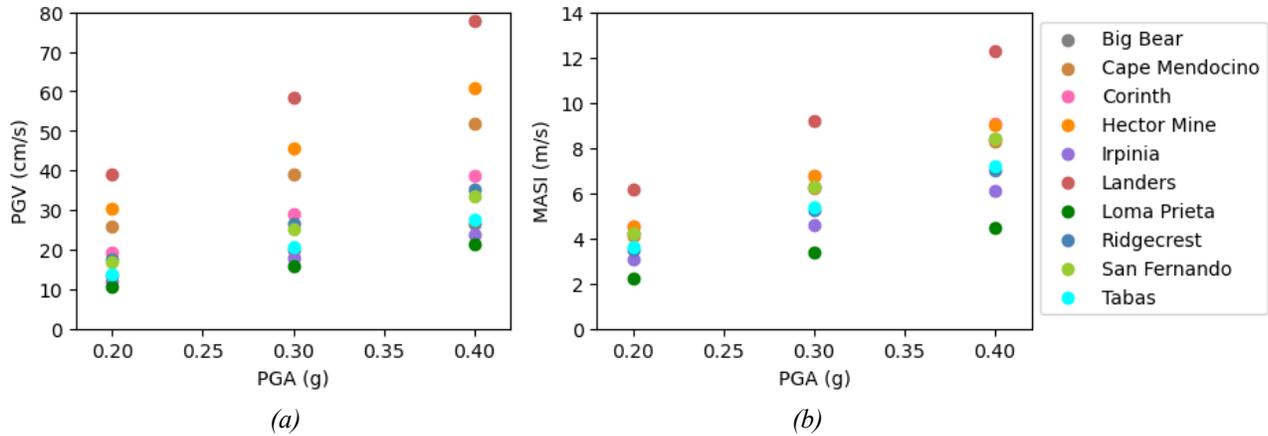

*Figure 5: (a) PGV vs. PGA and (b) MASI vs. PGA for the input ground motions.*

## INFLUENCE OF PGA ON LIQUEFACTION FAILURES

The influence of ground motion PGA on the Mochikoshi tailings dam failure is investigated by considering the Ridgecrest ground motion scaled to PGA levels of 0.2, 0.3, and 0.4 g. These scaled Ridgecrest ground motions are identical in frequency content and duration, varying only in PGA, allowing for a direct observation of how increasing PGA influences the extent of liquefaction and associated failure mechanism.

Figure 6a shows the liquefaction response in the FEM phase for these analyses in terms of depth of liquefaction as a function of time. The depth of liquefaction is defined as the elevation of the deepest node with $r_u > 0.7$ relative to the original pond surface elevation (y = 30 m). The depth of liquefaction is considered within two regions: below the crest of the dam (between x = -20 m and -40 m in Figure 1) and within the tailings pond (between x = -50 m and -70 m). For all three PGAs, the initial shaking ($t < 17$ s) induces liquefaction only at shallow depths (Figure 6a), but the liquefaction extends much deeper between $t \sim 17$ s and $t \sim 19$ s, the period of most intense shaking of the ground motion (Figure 4a). The maximum depths of liquefaction are achieved by $t \sim 20$ s, when the main shaking is over. As expected, when the PGA is more intense, the maximum depth of liquefaction is greater. The motion scaled to PGA = 0.2 g only induces liquefaction to a depth of ~5 to 8 m, while liquefaction extends to ~16 to 18 m for PGA = 0.3 g and ~23 to 28 m for PGA = 0.4 g. The depths of liquefaction in the pond and below the crest show similar trends, accumulating to similar depths at similar times, but differences in confining pressures and static shear stresses near the dam result in some differences.

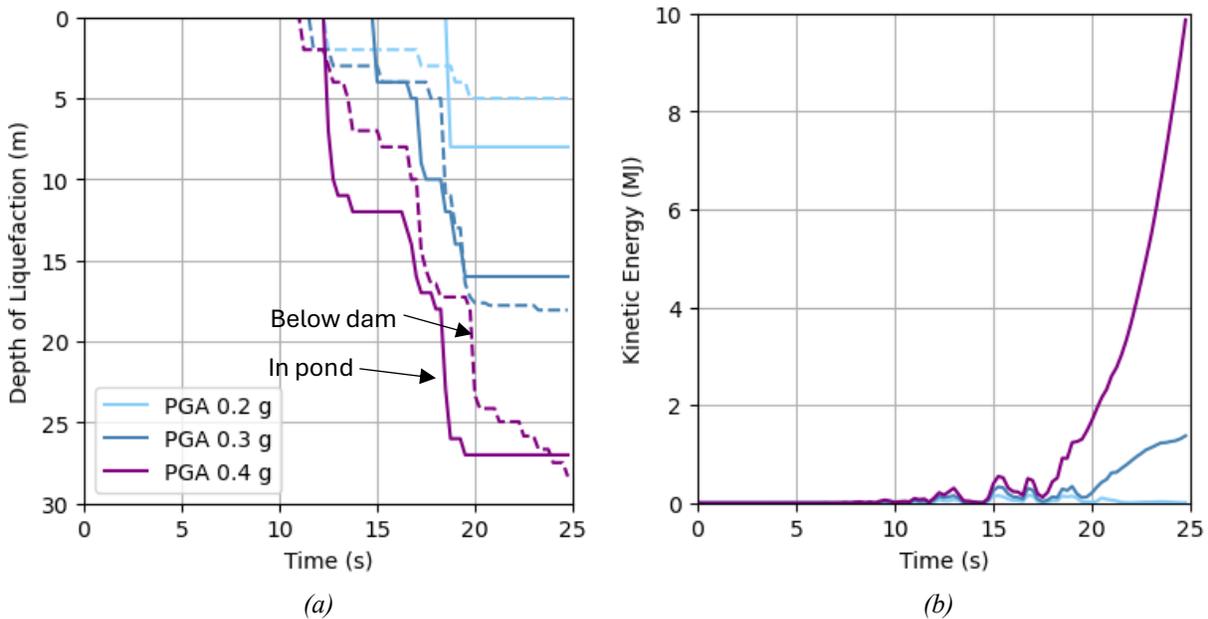

*(a)*  *(b)*

*Figure 6: (a) Depth of liquefaction and (b) kinetic energy vs. time resulting from Ridgecrest ground motions at PGAs of 0.2, 0.3, and 0.4 g.*

Figure 6b plots the kinetic energy of the FEM phases as a function of time for the three analyses. The kinetic energy of the FEM phase is defined as the sum of the kinetic energy of every element, computed from the element mass and the average velocity of the four nodes of the element [21]. The analyses for PGA of 0.3 g and 0.4 g show abrupt increases in kinetic energy at approximately $t$ = 19.5 s and 17.5 s, respectively (Figure 6b), and these times are associated with liquefaction developing to a depth of about 16 to 18 m (Figure 6a). These liquefaction depths associated with increasing kinetic energy are consistent with limit equilibrium analyses that indicate a factor of safety (FS) of 1.0 for a liquefaction depth of 12 m and a FS of 0.5 for liquefaction depths of 18.5 m and deeper [22]. The increase in kinetic energy is more rapid in the PGA = 0.4 g analysis because the liquefaction depth extends deeper, allowing a larger mass within the pond to mobilize. In contrast, the increase in kinetic energy in the PGA = 0.3g analysis occurs later because it takes more cycles to induce liquefaction to the critical depth, and it does not increase as rapidly because the resulting liquefaction depth is smaller. The PGA = 0.2g analysis does not exhibit any significant increase in kinetic energy, as the intensity of shaking is insufficient to cause liquefaction to the critical depth by the end of the ground motion.

The transfer time for each of these analyses was selected based on the guidelines described in our previous publication [22]. These guidelines state that the transfer should take place after liquefaction failure has fully developed and after the kinetic energy has begun to progressively increase (if sufficient liquefaction is reached) but before the mesh has become excessively distorted. Excessive distortion is defined as the time at which the average normalized Jacobian determinant $((|J_t| / |J_0|)_{avg})$ among elements with a deviatoric strain ($\varepsilon_q$) greater than 3% falls below 0.97. For the PGA = 0.3 and 0.4 g analyses, the transfer times were 23 s and 20.5 s, respectively, values that lie approximately halfway between full failure development and excessive mesh distortion. For the PGA = 0.2 g analysis, the transfer is performed at the end of the time history, because no increase in kinetic energy occurs and mesh distortion is negligible.

Figure 7 illustrates the deviatoric strain contours from the FEM analyses at the transfer time for each of the three input PGA values. These contours reveal how deeper liquefaction affects the failure mechanisms across different PGA values. The PGA = 0.2 g analysis shows no significant shearing, as the PGA is insufficient to cause substantial liquefaction. In contrast, the PGA = 0.3 g and 0.4 g analyses exhibit significant deviatoric strains, indicating large-scale failure mechanisms. The initial failure-inducing shear surface forms below the dam (x > -30 m) for both of these cases, and thus the liquefaction depth below the dam is associated with triggering the initial dam failure. While the shear surface below the dam is similar for both cases, its extension into the pond exhibits some variations. In the PGA = 0.4 g analysis, the extension of the shear surface further into the pond (x < -60 m) is approximately 4 m deeper (~ 11 m deep) compared to the PGA = 0.3 g analysis (~ 7 m deep). This deeper shear surface extending further into the pond allows a greater mass of material to be mobilized, resulting in more accumulated kinetic energy for the PGA = 0.4 g motion.

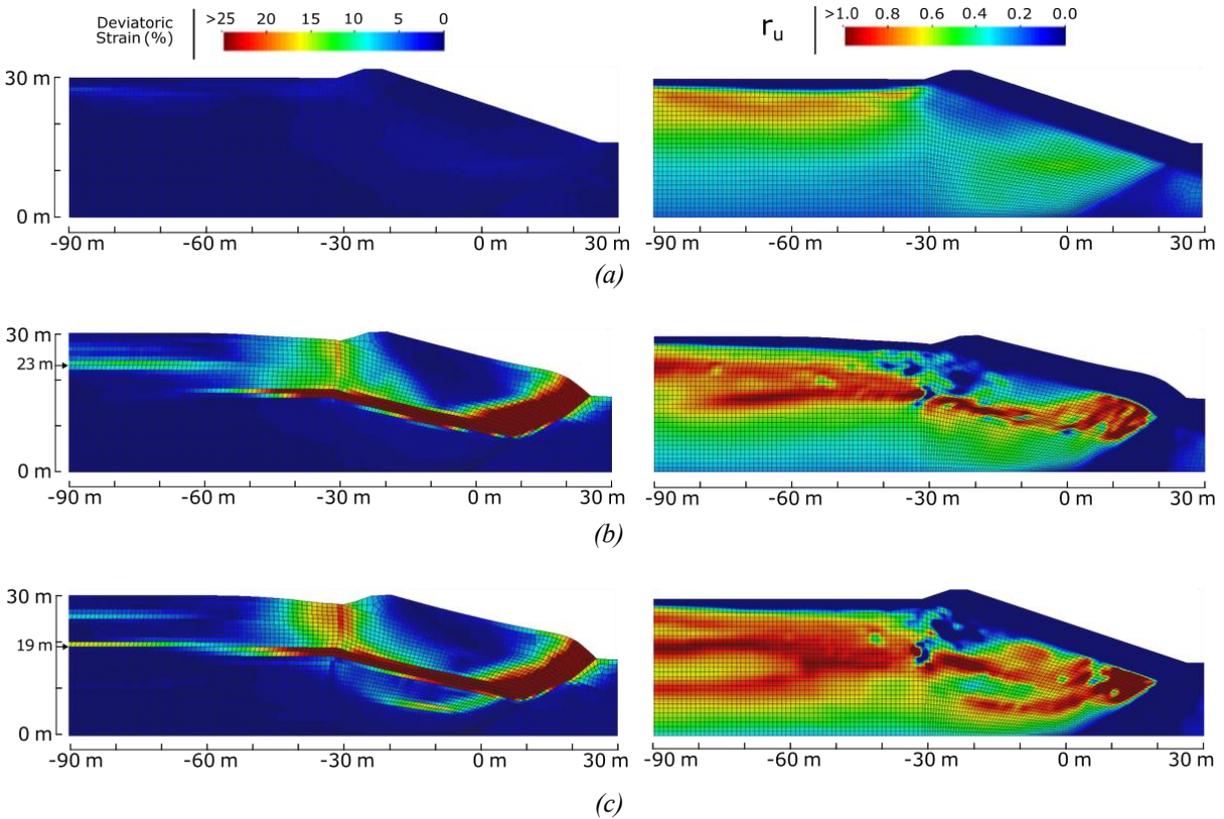

*Figure 7: Deviatoric strain and $r_u$ contours of the FEM phase at the transfer time for analysis using the Ridgecrest ground motion at PGAs of (a) 0.2, (b) 0.3, and (c) 0.4 g.*

Figure 8 illustrates the final runout geometries for the MPM analyses with $c_{soft} = 0.5 \cdot S_r$ for the three PGA values. The results demonstrate a clear relationship between PGA and the extent of dam failure and material runout. The PGA = 0.2 g analysis (Figure 8a) yields no runout, as the depth of liquefaction in the FEM phase was insufficient to cause a failure. Both the PGA = 0.3 and 0.4 g analyses (Figures 8b, c) result in failure of the upper dam, but with significantly different final runouts. The PGA = 0.4 g analysis leads to a substantial release of material and a runout distance of ~220 m, as compared with a runout distance of ~90 m for the PGA 0.3 g analysis. The displacement contours in Figure 8 reveal that tailings material at lower elevations (i.e., originating from greater depths in the pond) are mobilized and experience larger displacements in the PGA = 0.4 g analysis. In the PGA = 0.3 g analysis, the deepest mobilized material points are at y ~ 20 m (depth = 10 m), while in the PGA = 0.4 g analysis, they reach y ~ 13 m (depth 17 m). This observation aligns with the deeper liquefaction (Figure 6a) and deeper shear bands (Figure 7) from the FEM analysis for the PGA = 0.4 g analysis. Notably, the depths of mobilized material points are greater than the shear surfaces in their respective FEM analyses (Figure 6), indicating that particles continue to be mobilized as the runout evolves in the MPM phase.

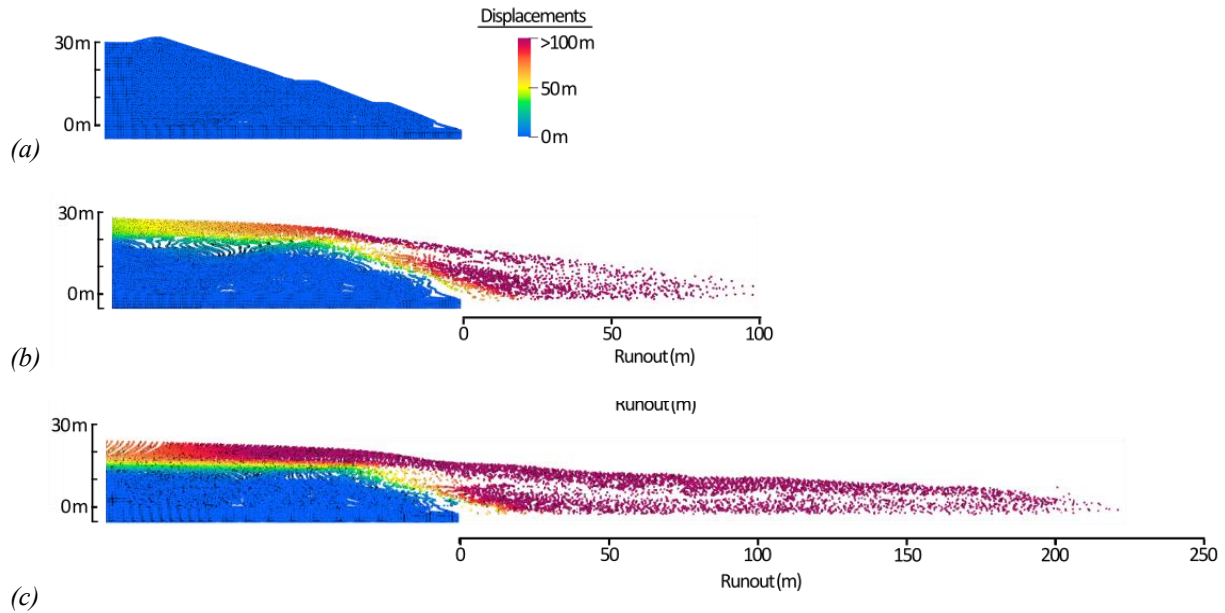

*Figure 8: Final runout geometries resulting from Ridgecrest ground motion at PGA of (a) 0.2, (b) 0.3, and (c) 0.4 g when $c_{soft} = 0.5 \cdot S_r$.*

**INFLUENCE OF GROUND MOTION CHARACTERISTIS ON LIQUEFACTION FAILURES**

Although traditional stress-based liquefaction analyses, as noted by NASEM [9], typically rely solely on PGA and earthquake magnitude to predict liquefaction, research has demonstrated that other ground motion characteristics play a significant role in the liquefaction process. For instance, Kramer and Mitchell [10], Nong et al. [36], and Qin et al. [37] have shown that factors such as frequency content and duration substantially affect both the rate of pore pressure generation and the maximum extent of liquefaction. The influence of these ground motion characteristics on liquefaction evolution and ultimate runout is explored by analyzing four different input motions: Loma Prieta, Ridgecrest, Hector Mine, and Landers, each scaled to a PGA of 0.3 g. By comparing the response of the tailings dam to these four ground motions, this study aims to provide a more comprehensive understanding of how various ground motion properties influence liquefaction development and subsequent tailings dam runout.

Figure 9 illustrates the temporal evolution of the depth of liquefaction and kinetic energy for the four different motions, tracked until the end of each time history or until mesh entanglement occurs. The Loma Prieta motion results in minimal liquefaction, reaching only ~5 m depth (Figure 9a) by the end of its time history.

Consequently, there is no increase in kinetic energy (Figure 9b), indicating insufficient liquefaction to trigger dam failure. The Ridgecrest ground motion induces liquefaction to a depth of 16 m and experiences an increase in kinetic energy starting at ~20 s. However, the kinetic energy begins to level off and does not increase rapidly. The Hector Mine and Landers motions produce the deepest liquefaction (26 m and 24 m, respectively; Figure 9a) and both experience rapid increases in kinetic energy that do not level off. However, the temporal evolution of liquefaction differs between these two motions. The Landers analysis shows a rapid increase in liquefaction depth from 0 m to its maximum of ~22 m over a relatively short period of time, while the Hector Mine anaysis demonstrates a more gradual increase, reaching its maximum depth of ~26 m over a longer period of time. These differences in liquefaction evolution directly correspond to the time-domain characteristics of these motions (Figure 4a). The Landers motion concentrates its main shaking over ~5 s, proving more efficient at generating liquefaction. In contrast, the Hector Mine motion spreads its main shaking over a longer period, resulting in less efficient liquefaction generation. However, due to its extended duration, the Hector Mine motion eventually achieves a similar depth of liquefaction. Notably, in both the Hector Mine and Landers analyses, once liquefaction exceeds a depth of ~16 to 18 m, kinetic energy increases rapidly. This observation suggests a critical liquefaction depth beyond which dam stability is significantly compromised, regardless of the specific ground motion characteristics.

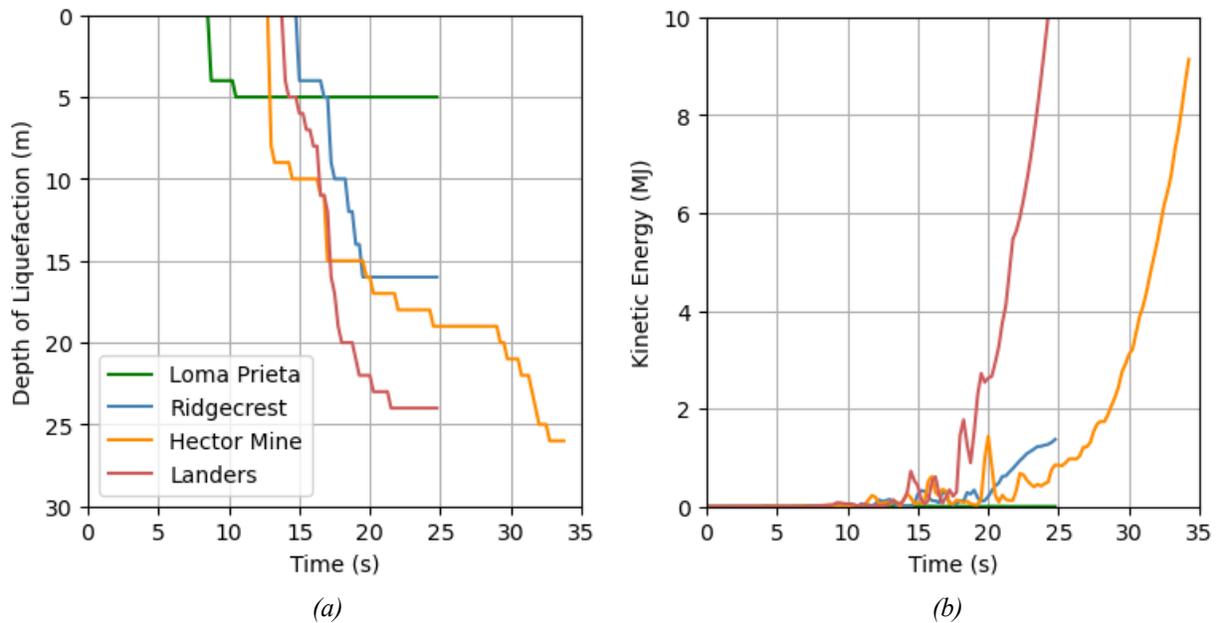

*(a)*                                    *(b)*

*Figure 9: (a) Depth of liquefaction in the pond and (b) kinetic energy as a function of time for selected ground motions at PGA = 0.3 g.*

Figure 10 illustrates the final runout geometries resulting from each analyzed ground motion, excluding Loma Prieta due to insufficient liquefaction-induced runout. The runout distances varied significantly: ~90 m for Ridgecrest and ~220 m for both Hector Mine and Landers. The smaller runout of the Ridgecrest analysis is attributed to its shallower maximum liquefaction depth (~16 m) compared to Hector Mine (~26 m) and Landers (~22 m). Notably, Hector Mine and Landers produced the same runout distance despite different liquefaction depths. This runout matches the runout observed in the Ridgecrest PGA = 0.4 g analysis (Figure 8), which had a maximum liquefaction depth of 23 m (Figure 6a). These results suggest that while deeper liquefaction generally increases runout, a threshold depth exists beyond which maximum runout is achieved. We previously [22] proposed this threshold at ~18.5 m based on limit equilibrium analysis, which aligns with these findings. Liquefaction depths below this threshold result in smaller failures with less runout (e.g., Ridgecrest PGA = 0.3 g) or no runout if sufficiently shallow (e.g., Loma Prieta PGA = 0.3 g and Ridgecrest PGA = 0.2 g). These observations highlight the complex relationship between liquefaction depth and runout distance in tailings dam failures, emphasizing the need to consider both the occurrence and extent of liquefaction when assessing seismic risks to tailings dams.

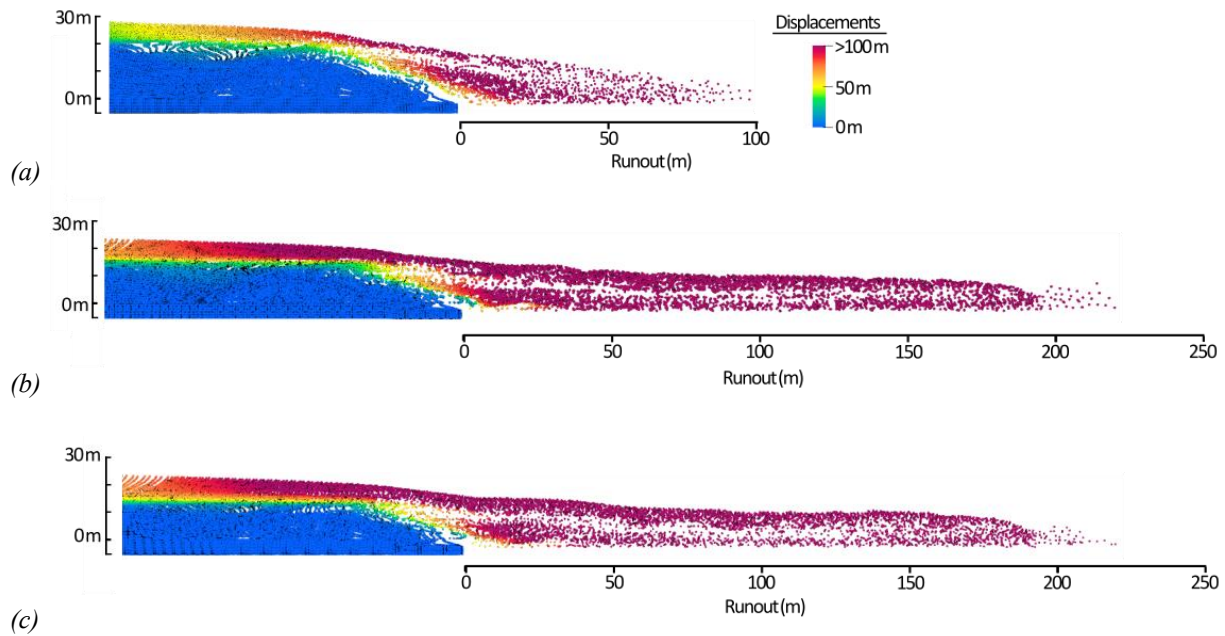

*Figure 10: Final runout geometries resulting from (a) Ridgecrest, (b) Hector Mine, and (c) Landers ground motions at PGA = 0.3 g with $c_{soft} = 0.5 \cdot S_r$.*

The different liquefaction and runout responses for ground motions with the same PGA indicate the influence of ground motion characteristics beyond PGA. It is notable that the $T_m$ of the Loma Prieta ground ($T_m = 0.22$ s, Table

4) is the smallest of all ten considered, while the $T_m$ of the Ridgecrest ground motion ($T_m = 0.39$ s) is intermediate, and the $T_m$ values of the Hector Mine and Landers ground motions ($T_m = 0.77$ and $0.66$ s, respectively) are among the largest. These mean periods are consistent with the response spectra (Figure 4b), where the Loma Prieta ground motion exhibits the weakest response at periods > 1 s, the Hector Mine and Landers ground motions have the strongest response, and the Ridgecrest ground motion is intermediate. This correlation suggests that lower frequency energy plays a significant role in inducing liquefaction in these models. Such findings are consistent with the experimental results reported by Nong et al. [36] and Qin et al. [37], further emphasizing the importance of considering frequency content, in addition to PGA, when assessing liquefaction potential and its consequences for tailings dam stability.

**INFLUENCE OF RESIDUAL STRENGTH ON RUNOUT**

The characterization of post-liquefied residual strength of the liquefied tailings in the MPM phase is a crucial parameter in runout prediction. This consideration is particularly important because phenomena like void ratio redistribution, as described by Kokusho and Kojima (2002), can alter material strength even after initial liquefaction. Previous analyses [22] explored two scenarios: one where the strength of the liquefied tailings remained constant ($c = S_r$) in the MPM phase of the analysis and another where the material strain-softened further to half of the post-liquefaction residual strength ($c_{soft} = 0.5 \cdot S_r$). To further investigate the sensitivity of runout results to residual strength, two additional cases are now analyzed: strain-softening to 75% and 25% of the post-liquefaction residual strength (i.e., $c_{soft} = 0.75 \cdot S_r$ and $c_{soft} = 0.25 \cdot S_r$).

Figure 11 illustrates the relationship between runout distances and the softened strength ratio (i.e., $c_{soft}/S_r$) for analyses using the four ground motions previously discussed. The results demonstrate a strong sensitivity to strain-softening, with runout distances increasing substantially as strain-softening intensifies. The Loma Prieta motion consistently yields no runout, regardless of the degree of strain-softening, due to insufficient liquefaction in the FEM phase to initiate failure (Figure 9). While the Ridgecrest, Hector Mine, and Landers motions all induce sufficient liquefaction for dam failure in the FEM phase, their runouts in the MPM phase differ. The Ridgecrest motion produces no runout without strain-softening ($c_{soft}/S_r = 1.0$, Figure 11), as liquefied material only flows partially down the dam face without reaching the toe. However, strain softening of $c_{soft}/S_r \leq 0.75$ enables the liquefied material to travel beyond the toe of the dam, although these runout distances remain consistently smaller than those of the Hector

Mine and Landers motions due to shallower liquefaction. The runout distances for the Hector Mine and Landers motions are consistently similar across all $c_{soft}/S_r$ because both motions induce liquefaction beyond the critical depth of $\sim 18.5$ m. These findings indicate that while strain-softening alone does not trigger runout failure (i.e., liquefaction must still reach a critical depth for failure mobilization), it significantly influences the travel distance of liquefied material. Consequently, when failure occurs, increasing levels of strain-softening lead to a significant increase in runout distance, underscoring the complex interplay between liquefaction depth and post-liquefaction material behavior in determining the extent of tailings dam failures.

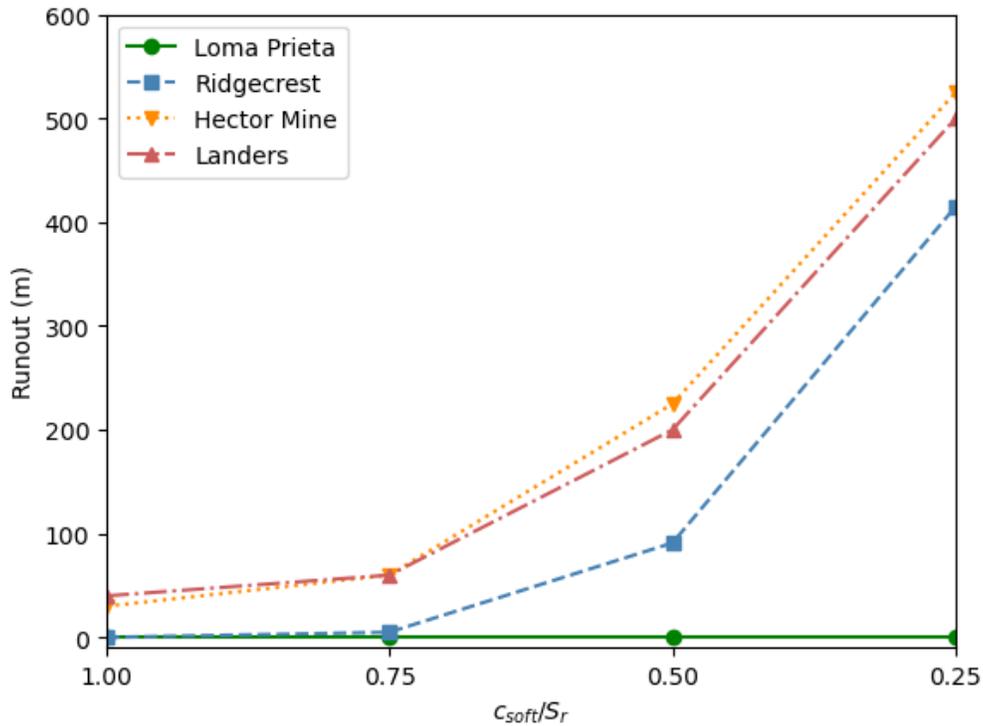

*Figure 11: Runout distances of hybrid models resulting from the Loma Prieta, Ridgecrest, Hector Mine, and Landers ground motions at PGA = 0.3 g.*

**COMBINED FACTORS INFLUENCING LIQUEFACTION TRIGGERING AND RUNOUT**

The depth of liquefaction induced by a specific ground motion is crucial in determining dam breach and runout occurrence. The results presented in the previous sections indicate that ground motions with identical PGA can generate varying liquefaction depths due to the effects of other ground motion characteristics, such as frequency content. To address this issue, additional ground motion Intensity Measures (IMs) are investigated to identify which parameters correlate most strongly with both the depth of liquefaction and the ultimate runout distance.

Figure 12 illustrates the maximum liquefaction depths generated by each of the ten ground motions at the three PGA values. At PGA = 0.2 g, liquefaction depths range from 3 to 14 m. For PGA = 0.3 and 0.4 g, the depths are larger, with most motions causing liquefaction deeper than 15 m, but significant variations exist among the motions. At each PGA, smaller liquefaction depths generally correlate with smaller $T_m$ (< 0.4 s) and larger depths with larger $T_m$ (> 0.5 s). As noted earlier, this demonstrates that both PGA and frequency content (i.e., $T_m$) influence liquefaction depth. Predicting liquefaction development based solely on PGA overlooks a crucial ground motion characteristic affecting pore pressure development.

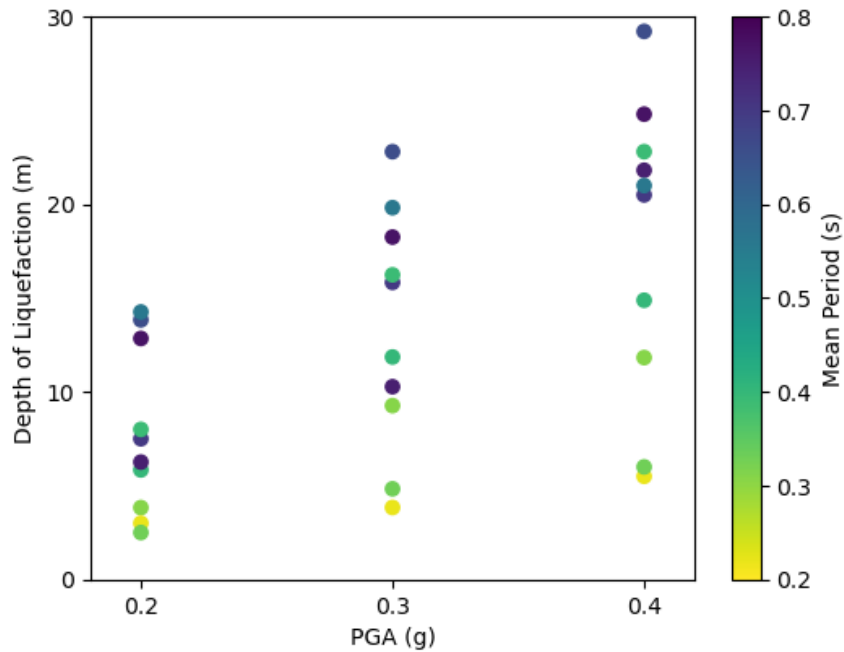

*Figure 12: Depth of liquefaction in pond as a function of PGA and mean period.*

To better represent this phenomenon, a single IM that combines both intensity and frequency content is preferable to using two separate ground motion parameters. This study considers the ground motion parameters PGV and Maximum Acceleration Spectrum Intensity (MASI) for this purpose. Figure 13 illustrates the relationship between liquefaction depth in the pond and these two IMs. Both show a clear trend of increasing depth of liquefaction with increasing PGV and MASI, with significantly stronger correlations than those observed for PGA in Figure 12. Ground motions with PGV < 20 cm/s and MASI < 4 m/s generally produce liquefaction depths smaller than 10 m, and ground motions with PGV > 40 cm/s and MASI > 8 m/s generally produce liquefaction depths larger than 20 m. Ground motions with intermediate values of PGV or MASI yield intermediate liquefaction depths, but there is scatter in the data with liquefaction depths varying by about 15 m (i.e., liquefaction depths range from 5 to 20 m at PGV ~ 30 cm/s and MASI ~6 m/s). This scatter is potentially a result of PGV and MASI not fully accounting for the influence of frequency content, or a result of liquefaction depth being influenced by other ground motion characteristics, such as duration.

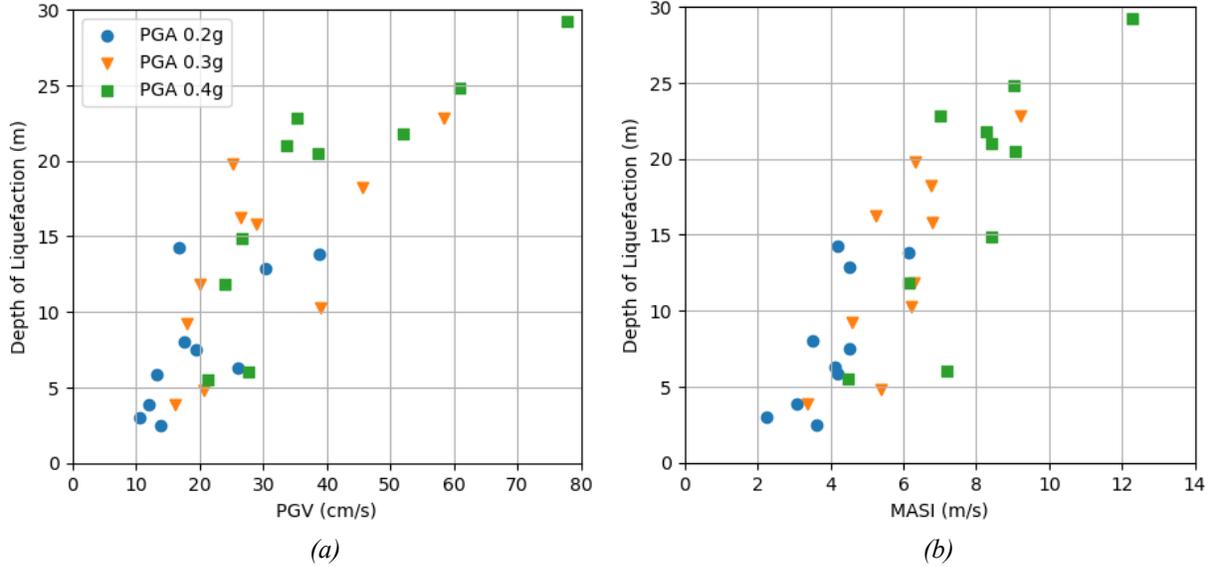

*Figure 13: Depth of liquefaction in the pond as a function of (a) PGV and (b) MASI*

Figure 14 shows the final runout distances from each ground motion as a function of the depth of liquefaction in the pond and the softened strength ratio ($c_{soft}$ /$S_r$) of the liquefied materials. The analyses consistently reveal a clear relationship between deeper liquefaction and larger runout distances. A liquefaction depth of approximately 10 m appears to be the critical threshold for dam failure and measurable runout, with ground motions producing liquefaction depths less than 10 m consistently resulting in no runout. As the depth of liquefaction increases beyond this threshold, the runout distance also increases. This trend continues until a second threshold is reached at ~18 m of liquefaction depth, where runout distances become relatively consistent for larger liquefaction depths. For the tailings dam geometry analyzed in this study, liquefaction depths beyond ~ 18 m do not result in additional runout because the lower dam is thick enough to remain intact and contain the deepest tailings even if they liquefy. This is concurrent with our previous findings [22]. These results suggest that while liquefaction depth significantly influences runout, other factors play a more dominant role in determining runout distances beyond a critical liquefaction depth (i.e., 18 m). This behavior is specific to this tailings dam design and will be different for other geometries.

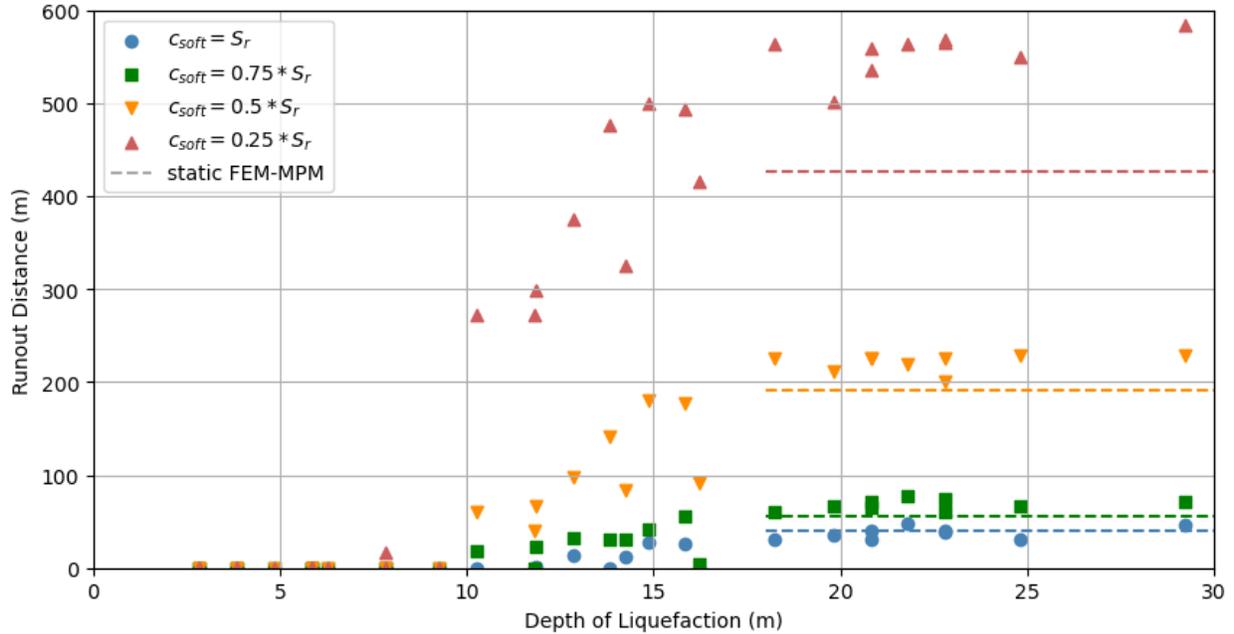

*Figure 14: Runout distance for different softened strength ratios as a function of depth of liquefaction.*

Another important observation from Figure 14 is that the degree of softening of the liquefied soil strongly influences runout. For $c_{soft}$ /$S_r$ = 1.0 or 0.75, which represents undrained residual strengths within 25% of common empirical correlations [14,29], moderate runouts of only 50-75 m are obtained. However, runout distances of hundreds of meters, which are more consistent with field observations of tailings dam failures, are computed only when smaller softened strengths are used (i.e., $c_{soft}$ /$S_r$ = 0.5 or 0.25). Thus, the assignment of the softened strength of the liquefied materials is a critical component of the analysis when predicting runout distances.

For a final comparison of the hybrid FEM-MPM procedure to alternative methods of estimating runout of liquefaction-induced flow slides, the hybrid models are compared to a static hybrid FEM-MPM analysis that neglects the dynamic effects of ground shaking. The static FEM-MPM model features the same tailings dam geometry and material properties, but it only receives the geo-static state of stress from the FEM phase, and it assumes that all of the tailings liquefy. This is a model that would be compliant with ANCOLD [39] guidelines, which recommends assuming all liquefiable materials liquefy for stability analysis. This assumption is the industry standard for static liquefaction analysis [40] and could be considered conservative for an analysis of a seismic failure.

The runout results from the static FEM-MPM analyses are shown in Figure 14. Despite being ostensibly conservative by assuming liquefaction of all of the tailings, the static FEM-MPM analysis generally predicts less runout than the dynamic FEM-MPM analysis for which the depth of liquefaction exceeds 18 m. For $c_{soft}$ /$S_r$ = 1.0, the runout of 40 m from the static FEM-MPM analysis falls within the range of runouts from the dynamic FEM-MPM analyses. However, for $c_{soft}/S_r < 1.0$, the runout predictions from the static FEM-MPM analyses are 15-20% below the average of the dynamic FEM-MPM analyses. These results indicate that the development of the initial failure surface in the dynamic FEM phase of the analysis contributes to the final runout prediction. Given that the differences between the static and dynamic FEM-MPM analyses occur only when strain softening is included in the liquefied tailings, we attribute the difference to the deviatoric strains developed during the FEM phase of the analysis, specifically those resulting from the dynamic effects of the earthquake. Neglecting these earthquake-induced deviatoric strains in the static FEM-MPM analysis delays the strain-softening effects until the static failure has developed enough deviatoric strains to reduce the strength, resulting in a noticeable reduction in runout.

**CONCLUSIONS**

This study employs a sequential hybrid FEM and MPM analysis to assess the sensitivity of tailings dam failure to the characteristics of earthquake ground shaking and to the characterization of the post-liquefaction residual strength of the tailings. Using the geometry of the Mochikoshi tailings dam, which failed due to earthquake-induced liquefaction, dynamic FEM analyses were performed for 30 different earthquake motions and each FEM analysis was transferred to MPM to compute runout. The transfer process involves transferring the deformed geometry, state variables, and material properties from FEM to MPM, including the extent of liquefaction. In the MPM analysis, the liquefied materials are assigned post-liquefaction residual strengths and some analyses considered further softening of this residual strength.

The analyses considered ten earthquake acceleration-time histories, each scaled to PGA values of 0.2, 0.3, and 0.4 g, totaling 30 ground motions. While higher PGAs predictably generate deeper liquefaction and greater runout distances, the results demonstrate that different ground motions with the same PGA can produce significantly different liquefaction depths. Ground motion characteristics beyond PGA, particularly frequency content, play a crucial role in pore pressure generation and liquefaction depth. Ground motions with more long-period energy, as measured by the mean period, generate liquefaction to larger depths. Ground motion characteristics that account for both intensity and frequency content (i.e., peak ground velocity (PGV) and Modified Acceleration Spectrum Intensity (MASI)) correlate more strongly with liquefaction depth than PGA.

For the analyzed tailings dam geometry, a liquefaction depth of approximately 10 m in the tailings pond is necessary for measurable runout, with runout increasing with increasing liquefaction depths. The maximum runout occurs for liquefaction depths greater than about 18 m. Liquefaction beyond 18 m depth does not further affect runout for the geometry analyzed due to the lower portion of the embankment dam remaining intact. The runout distance proves highly sensitive to the post-liquefaction residual strength characterization of the tailings, with additional strain-softening during runout leading to significant increases in runout distance.

Also considered is a static FEM-MPM analysis that transfers a static stress state from FEM to MPM and simply assigns liquefied residual strengths to all tailings. When further strain-softening of the liquefied residual strength is considered, the static FEM-MPM analyses predict 15 to 20% lower runout distances compared to the dynamic FEM-MPM analyses. This difference arises from the dynamic FEM-MPM analyses capturing earthquake-induced deviatoric strains, resulting in more rapid strain softening in the MPM phase of the analysis. When no additional strain-softening is modeled in the tailings, the static and dynamic FEM-MPM analyses provide similar runout results.

This study underscores the importance of selecting appropriate ground motions when evaluating the seismic response of tailings dams, as both the intensity and frequency content of earthquake shaking significantly influence the depth and extent of liquefaction and subsequent dam failure and runout. Furthermore, the strength of liquefied tailings during the runout phase substantially affects runout distance, highlighting the need for additional research to better characterize the appropriate strength of liquefied materials during flow failures.